# Experimental Demonstration of High-Precision Multi-access Time Transfer via Optical-Electrical-Optical Repeater Stations

Hao Zhang[1], Guiling Wu[1,2,*], Xinwan Li[1] and Jianping Chen[1,2]

[1]*State Key Laboratory of Advanced Optical Communication Systems and Networks, Department of Electronic Engineering, Shanghai Jiao Tong University, Shanghai 200240, China*
[2] *Shanghai Key Laboratory of Navigation and Location Based Services, Shanghai 200240, China*
* *wuguiling@sjtu.edu.cn*

**Abstract:** In this paper, we propose a kind of optical-electrical-optical (OEO) repeater station for the time transfer based on bidirectional time division multiplexing transmission over a single fiber with the same wavelength (BTDM-SFSW). Along the main fiber link, the forward and backward time signals alternatively pass through the same OEO repeater via a 2×2 optical switch (OS). By this means, the optical powers of bidirectional time signals increase. And parts of excess noises, suffered from the main link, are sufficiently blocked owing to the OEO process. Simultaneously, the bidirectional time signals can be decoded and utilized for multi-access time transfer. As a result of the same devices for both directions, the maximum delay symmetry is maintained, and the performance for multi-access time transfer is guaranteed. With the OEO repeater stations, we successfully demonstrate multi-access time transfer in our time transfer testbeds with the main links up to 13,200 km. The time deviations of less than 80 ps/s and 11 ps/$10^5$ s, respectively, are obtained. Furthermore, the combined standard uncertainties of less than 70 ps are reached, which agree well with the experimental verifications over non-calibrated fiber links.

## 1. Introduction

In the past decade, fiber-optic has proven itself as one of the most promising tools for applications requiring highest precision in time metrology [1]. Besides point-to-point ones, a variety of multi-access fiber-optic time transfer have been demonstrated. For multi-access, a simple solution is multiple sets of point-to-point time transfer equipment incorporating with each other [2-5]. However, the overall hardware and system control are complex when the end users are large. The other method is tapping the counter-propagating time signals at any access point (AP) along the main fiber link [6-8]. The key point [9] is determining the time interval between bidirectional time signals at each AP. Nevertheless, bidirectional optical amplifiers are imperative as the total signal extraction is great. Once the bidirectional optical amplifier are installed along the main link, the undesired noises, such as double Rayleigh backscattering (DRB) travelling around one or more amplifiers [10], and the single Rayleigh backscattering (SRB) of backward amplified spontaneous emission (ASE) noise, etc., will significantly degrade the signal-to-noise ratio (SNR) of the received time signals. Subsequently, the total fiber link extension has to be limited for an acceptable time transfer precision [10, 11]. In addition, the time signals back and forth go through different paths at AP, which deteriorates the stability and uncertainty of time transfer [6, 7].

Previously we proposed a time transfer scheme using bidirectional time division multiplexing transmission over a single fiber with the same wavelength (BTDM-SFSW) [12]. In this paper, we propose an optical-electrical-optical (OEO) repeater station for BTDM-SFSW based time transfer. Via a 2×2 optical switch (OS), the forward and backward time signals along the main link alternatively pass through the same OEO repeater. By this means, the bidirectional time signals are amplified equivalently. And the excess noises suffered from the main link are sufficiently blocked. Simultaneously, the bidirectional time signals can be decoded and utilized for multi-access time transfer. As a result of the same devices for both directions, the maximum delay symmetry is maintained and the performance for multi-access time transfer is guaranteed. Via the OEO repeater stations, we successfully demonstrate multi-access time transfer in our time transfer testbeds with the main links up to 13,200 km. The time deviations of less than 80 ps/s and 11 ps/$10^5$ s, respectively, are obtained. Without fiber link calibration, the combined uncertainties of less than 70 ps are reached.

## 2. Principle

Fig. 1 summaries the structures of proposed OEO repeater stations. The basic configuration of OEO repeater is illustrated in Fig. 1 (a). For BTDM-SFSW based time transfer, the forward and backward optical time signals are carried on the same wavelength of λ. In different time slots, they arrive at each OEO repeater station, and then are sent into the conventional OEO repeater through the control of 2×2 OS. The newly generated time signals are launched into the corresponding fiber spans via the 2×2 OS. In this way, the bidirectional time signals are amplified equivalently. And the excess noises suffered from the main link, such as the DRB travelling around one or more optical amplifiers, and the SRB of backward ASE, etc., can be appreciably eliminated. Thus the SNR of the received time signals can be improved.

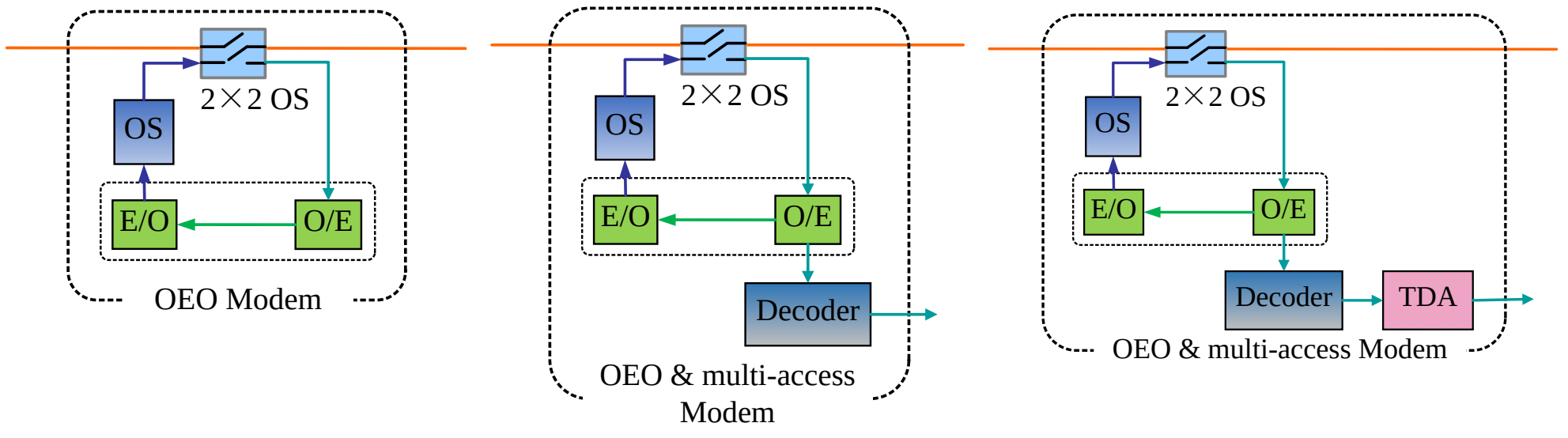

(a) (b) (c)

Fig. 1 Schematic structures of proposed optical-electrical-optic (OEO) repeater station. (a) basic configuration; (b) multi-access time transfer with local time scale; (c) multi-access time transfer without local time scale; TDA: time delay adjuster.

Simultaneously, the bidirectional time signals can be decoded and utilized for multi-access time transfer (see Fig. 1 (b) and (c)). It should be noted that, at the absence of local time scale, the AP can obtain a stable time signal output incorporating with a time delay adjuster (TDA). Fig. 2 gives the corresponding schematic principle of multi-access time transfer. From Fig. 2 (b), the clock difference between the local site A and M can be calculated as (1). From Fig. 2 (c), the time interval for TDA at site M is determined as (2).

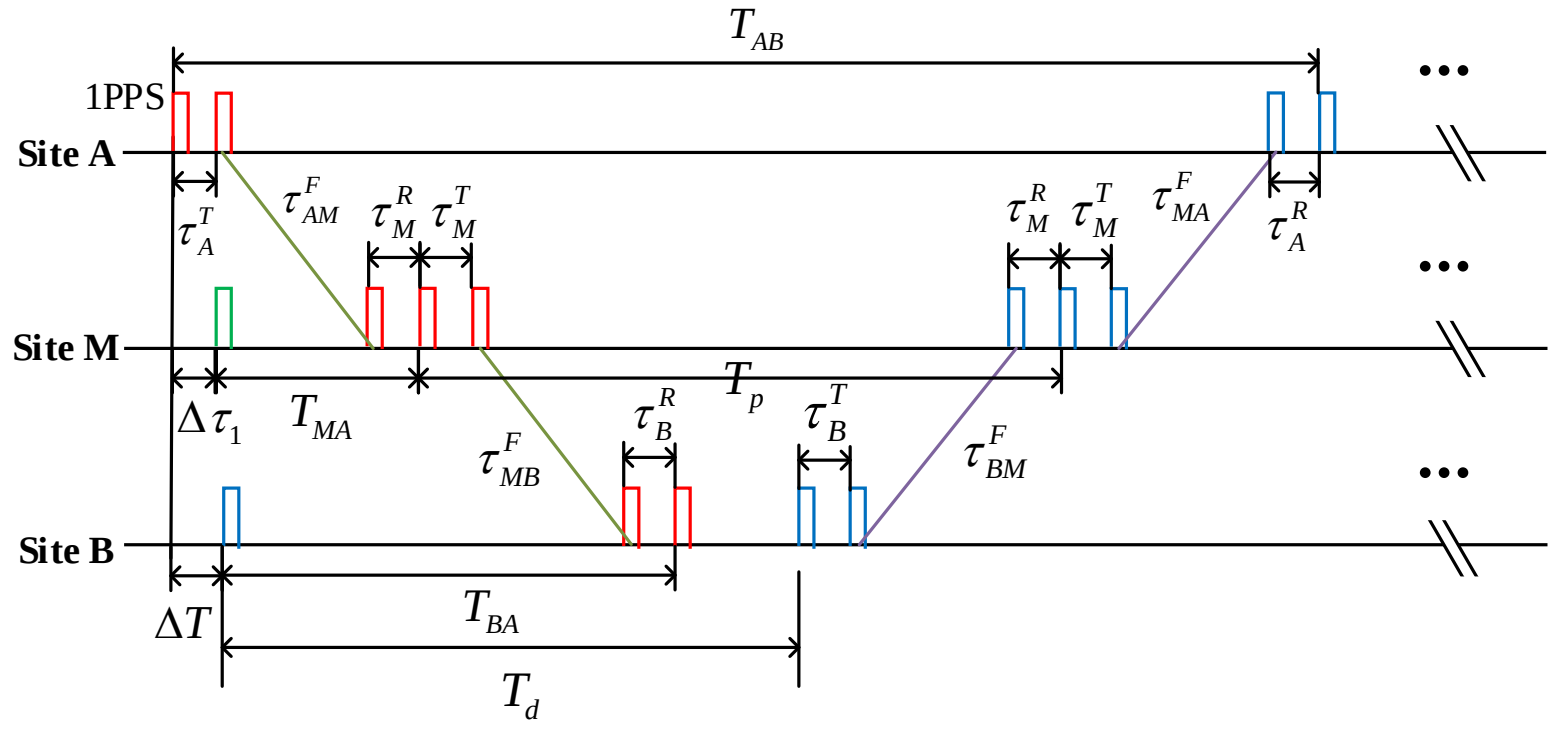

(a)

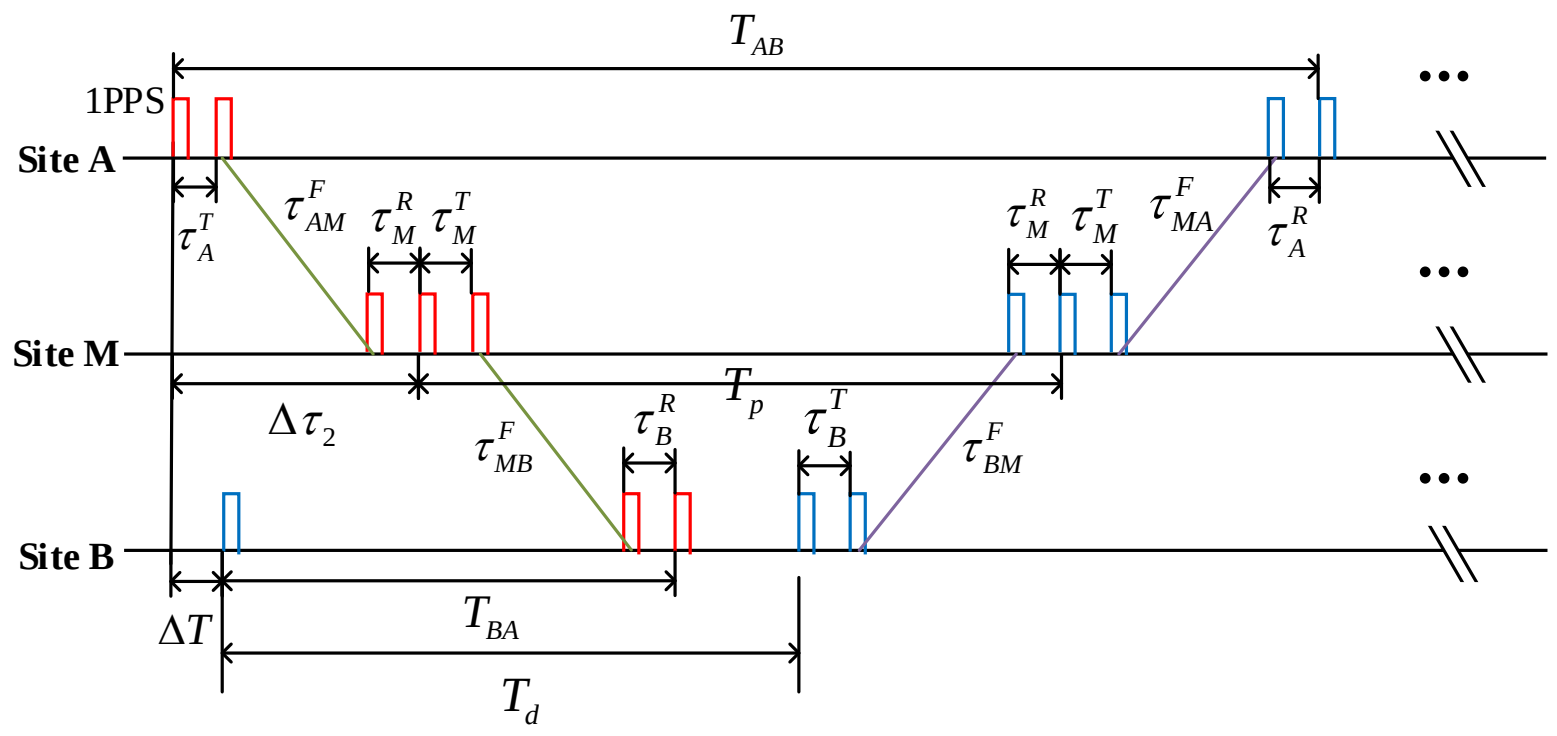

(b)

Fig. 2 Schematic principle of multi-access time transfer (a) with local time scale; (b) without local time scale.

$$\Delta\tau_1 = \frac{1}{2}\left[(T_{AB} - T_p - 2T_{MA}) + (\tau_A^T + \tau_M^R - \tau_M^T - \tau_A^R) + (\tau_{AM}^F - \tau_{MA}^F)\right]. \quad (1)$$

$$\Delta\tau_2 = \frac{1}{2}\left[(T_{AB} - T_p) + (\tau_A^T + \tau_M^R - \tau_M^T - \tau_A^R) + (\tau_{AM}^F - \tau_{MA}^F)\right]. \quad (2)$$

where $T_{AB}$ is the measured time difference between the local time signal and the received one at site A, $T_{MA}$ is the measured time difference between the local time signal and the one received at site M for the forward direction, $T_p$ is the measured time difference between the counter-propagating time signals received at site M, $\tau_{AM}^F(\tau_{MA}^F)$ is the propagation delay of fiber link from site A to site M (site M to site A), $\tau_A^T$ ($\tau_M^T$) and $\tau_A^R$ ($\tau_M^R$) are the sending and receiving delays at site A (site M), respectively.

As Fig. 1 shows, the optical time signals forth and back go through the same devices at each AP. The bidirectional delay difference within the 2×2 OS is negligible [13]. Therefore, the time transfer stability and uncertainty from repeater stations are guaranteed for multi-access. The bidirectional delay asymmetry, resulting from the wavelength difference, may increase with the repeater stations and be unacceptable. However, when the fiber spans connecting each repeater station is almost equal in length, the bidirectional delay asymmetry can be mitigated, even much better than the one of point-to-point link. For a proper operation of multi-access time transfer, the 2×2 OSs in each repeater stations should be set to the right states before the arrival of the corresponding time signals. And hence the link access control mechanism is required [13, 14].

## 3. Experiments and results

In order to evaluate the proposed scheme, a fiber-optic telecom testbed (see Fig. 3) is established. The whole system is placed in a normal air-conditioned laboratory. As shown in Fig. 3, site A and site B are connected by 350 km G.652 fiber, the corresponding dispersion compensated fiber (DCF), four signle-fiber bidirectional-transmission unidirectional optical amplifier (SFBT-UOAs) [13] and two OEO repeater stations. In the testbed, multi-access time transfer is implemented together with 10 Gb/s commercial data. Six WDMs with a channel spacing of 0.8 nm are employed in Fig. 3 to separate and combine the optical signals of time transfer and conventional network services. For time transfer, a common Rb clock (Symmetricom, 8040C) provides time signal (e.g. one pulse per second, 1 PPS) to site A and site B. Thus the clock drift on the test has no effect. With the corresponding OS, the optical signals at site A and site B are launched into the fiber link only during the transmission of the time codes. For multi-access, the time codes are modulated and received through commercially-available SFP transceivers at each repeater station. And the ITU-T wavelength channel is C35 (1549.32 nm). All the time intervals required are determined by the time interval counters (TICs,

Stanford Research System, SR620). For the conventional communication services, two optical carriers from the SFP transceivers with the wavelengths of $\lambda_2$=1550.12 nm and $\lambda_3$=1550.92 nm are modulated at 10 Gb/s in a non-return-to-zero (NRZ) modulation format with a $2^7$-1 pseudorandom binary sequence (PRBS). After a 170 km and 350 km transmission, respectively, the optical carriers are received by the transceivers and sent into the bit error rate tester (BERT) for BER evaluation.

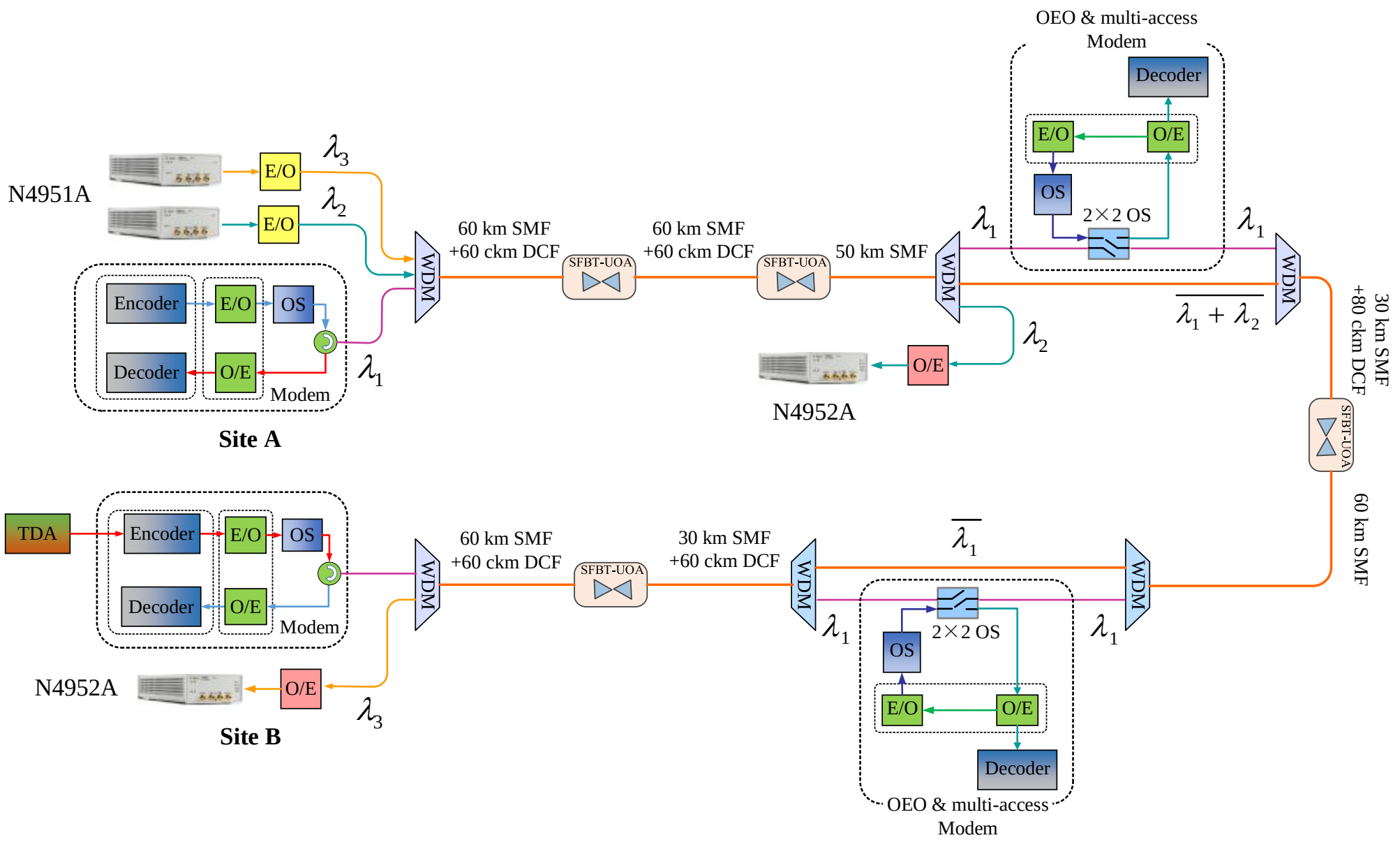

Fig. 3 Experimental fiber-optic telecom testbed.

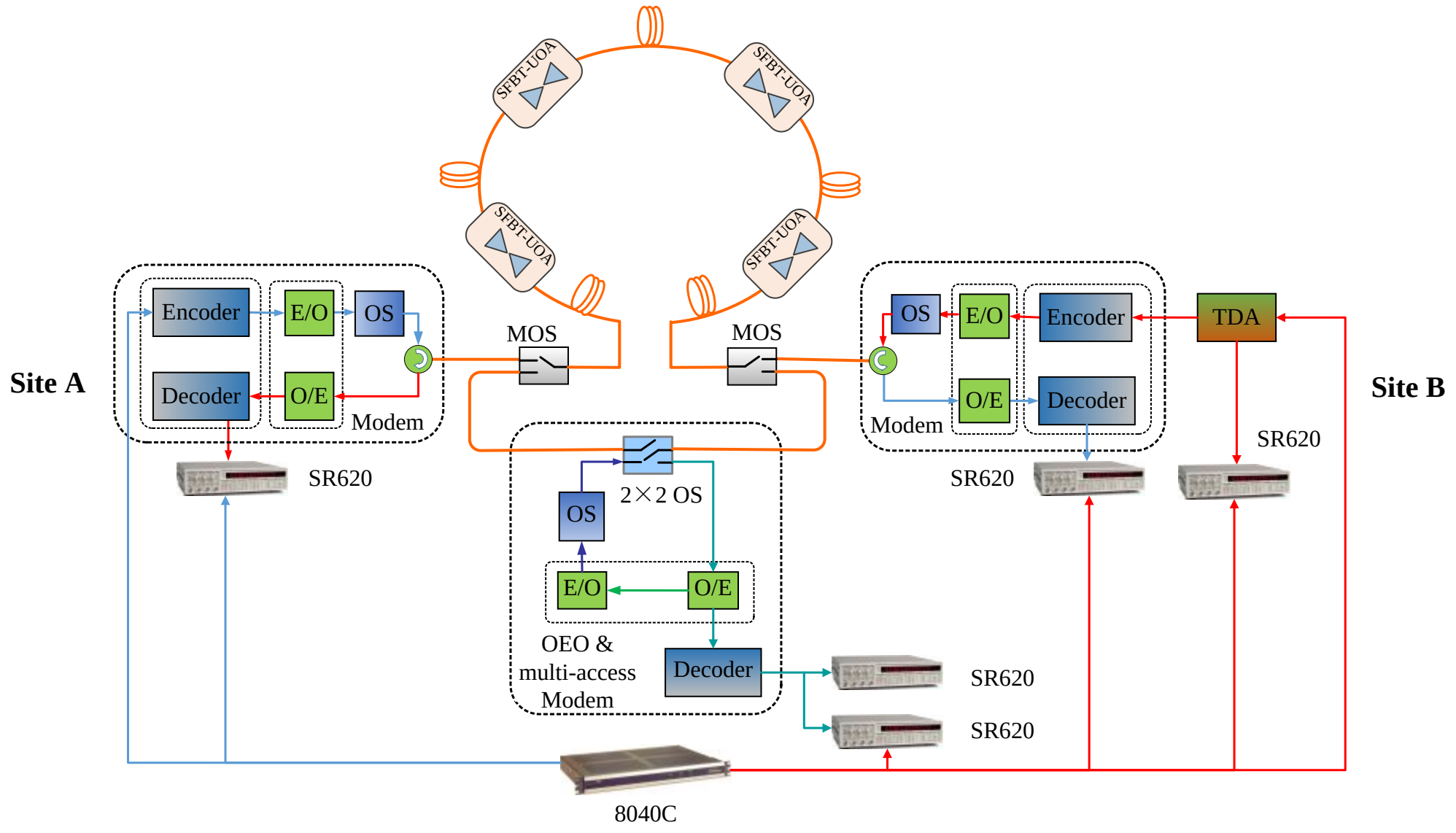

Fig. 4 Experimental setup of multi-access time transfer over ultra-long haul.

To further evaluate the performance, another experimental testbed of multi-access time transfer is established for ultra-long haul, as illustrated in Fig. 4. The optical signals are recirculated in the fiber loop, which is composed of 400 km G.652 fiber and four SFBT-UOAs.

The designated revolutions is controlled by two 1×2 magneto-optical switch (MOS). The MOSs have a maximum switching time of 50 μs.

Fig. 5 (a) plots the delay stabilities (time deviation, TDEV) of OEO repeater. A stability of less than 3 ps is observed for an averaging time of 1 s. And the TDEV drops down below 0.25 ps when the averaging time is larger than 1000 s. We can see that the effect of OEO repeater can be neglected for fiber-optic time transfer. Fig. 5 (b) gives the multi-access time transfer stabilities over 2 m and 350 km fiber links. Over the 2 m fiber link, the TDEVs for the main link amounts to 19.6 ps/s and 1.8 ps /$10^5$s. And the TDEVs of 21.9 ps/s and 2.3 ps /$10^5$s are observed for the AP. The stabilities of 2 m fiber link, regarding as the stability floor, are dominated by the adopted codecs, SFP transceivers and the TICs. The worse stability for AP can be attributed to the extra devices employed, such as TICs. As the link extends to 350 km, the TDEVs for the main link are deteriorated to 23.2 ps/s and 3.8 ps/$10^5$s. The TDEVs for the APs are worsen to 24.3 ps/s and 13 ps/$10^5$s, 23.9 ps/s and 2.9 ps/$10^5$s, respectively. The stabilities obtained in the 350 km fiber link are very similar.

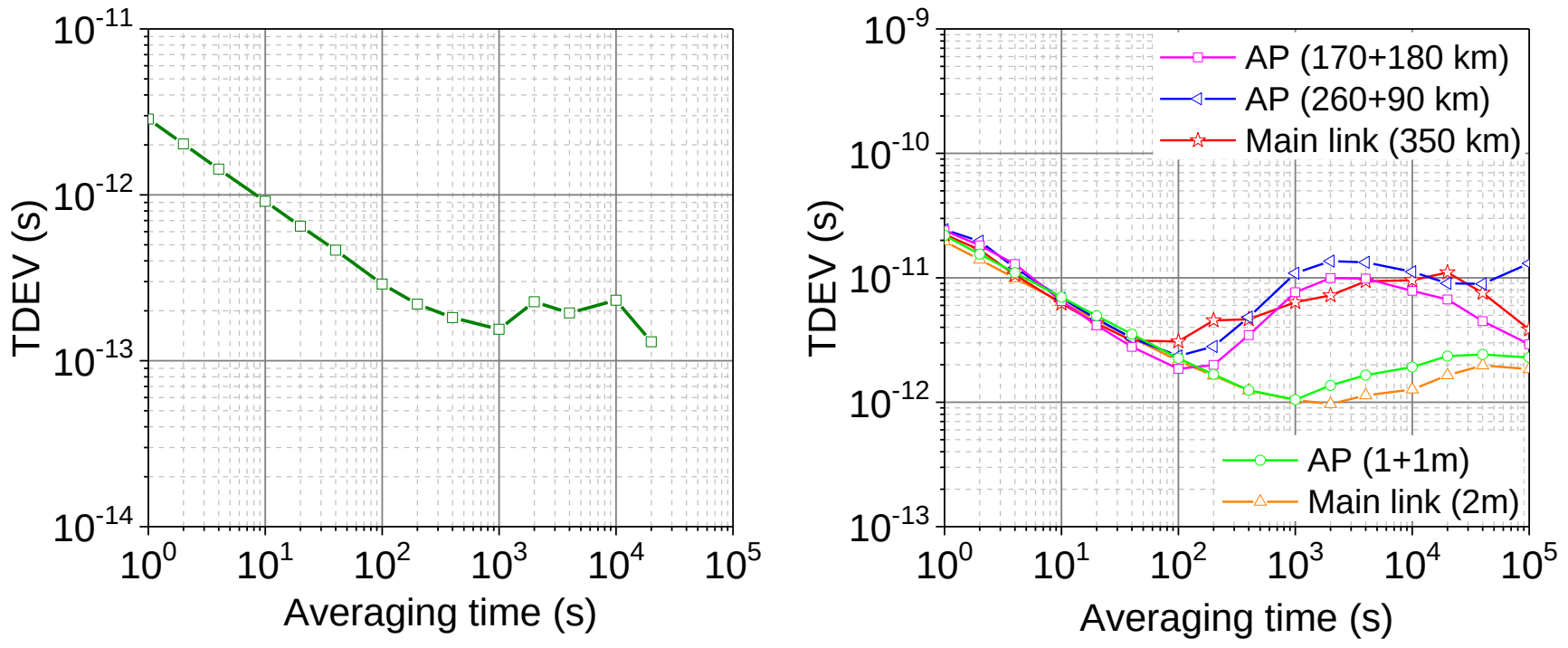

(a) (b)

Fig. 5 Time deviations (TDEVs) of (a) OEO repeater delay; (b) time transfer over 2 m and 320 km fiber links.

The measured BER for the 10 Gb/s data with a 170 km and a 350 km transmission, respectively, are given in Fig. 6. From the figure, the measured eye diagram can be clearly recognized. The BER is $8.3\times10^{-14}$ and $3.3\times10^{-13}$, respectively. The obtained results suggest that all the equipment adopted have the capability to support bidirectional time transfer and conventional network services at the same time.

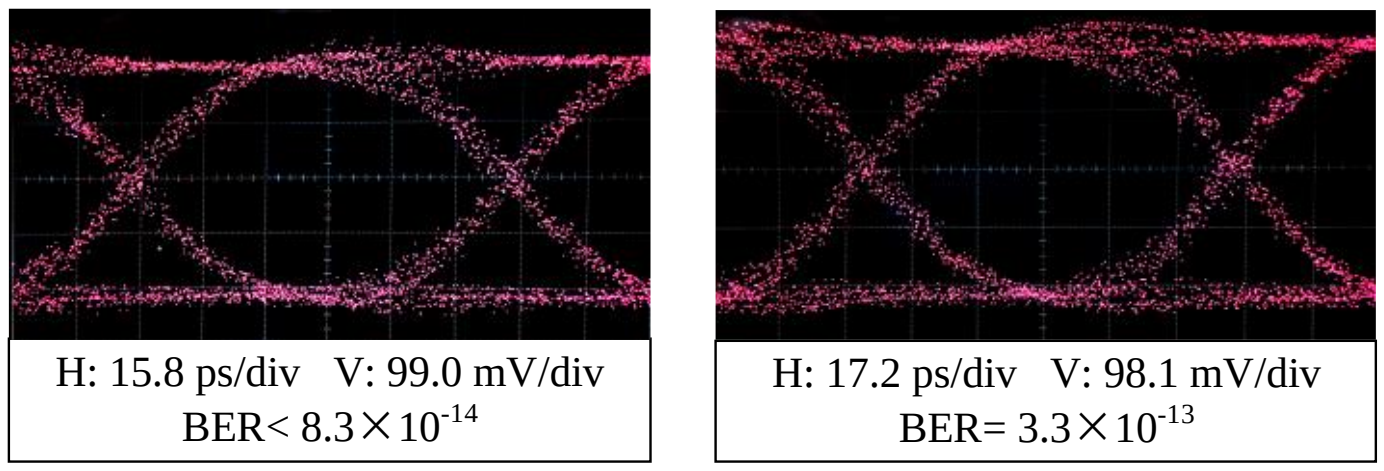

(a) (b)

Fig. 6 Measured BER for the 10 Gb/s data with the (a) 170 km and (b) 350 km transmissions.

Based on the ultra-long-haul experimental testbed, multi-access time transfer is demonstrated up to 13,200 km. Fig. 7 (a) plots the TDEV (1s) for multi-access time transfer. From the figure, the performance for the main link degrades as the fiber link extends from 400 km to 13,200 km. Compared with all optical fiber link [13], the receiving SNR improves significantly. Over the 13,200 km fiber link, the TDEVs (1 s) for any AP are less than 80 ps and have no evident trend. It may be explained that the required time intervals depend on each other and scale with the transmission length. Therefore, the total effect is nonlinear, which will be further analyzed in theory. Fig. 7 (b) shows the stabilities of multi-access time transfer over a 13,200 km fiber link. For the main link, the TDEVs are 53.7 ps/s and 10 ps /$10^5$s. For three APs, the stabilities approach the main link when the averaging time is below $10^4$s. And obvious difference is obtained as the averaging time increases further. Nevertheless, the TDEVs at the averaging time of $10^5$s are not larger than 11 ps yet.

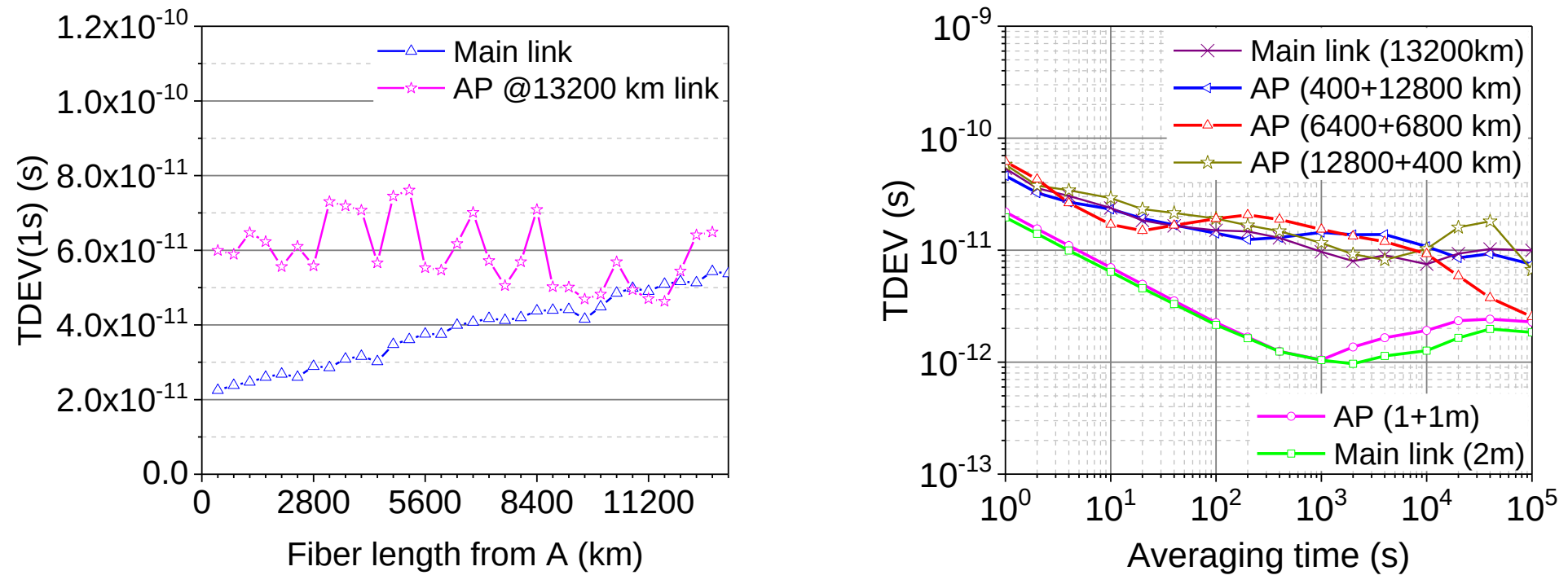

(a) (b)

Fig. 7. (a) TDEV (1s) for multi-access time transfer; (b) stabilities of multi-access time transfer over 2 m and 13,200 km fiber links.

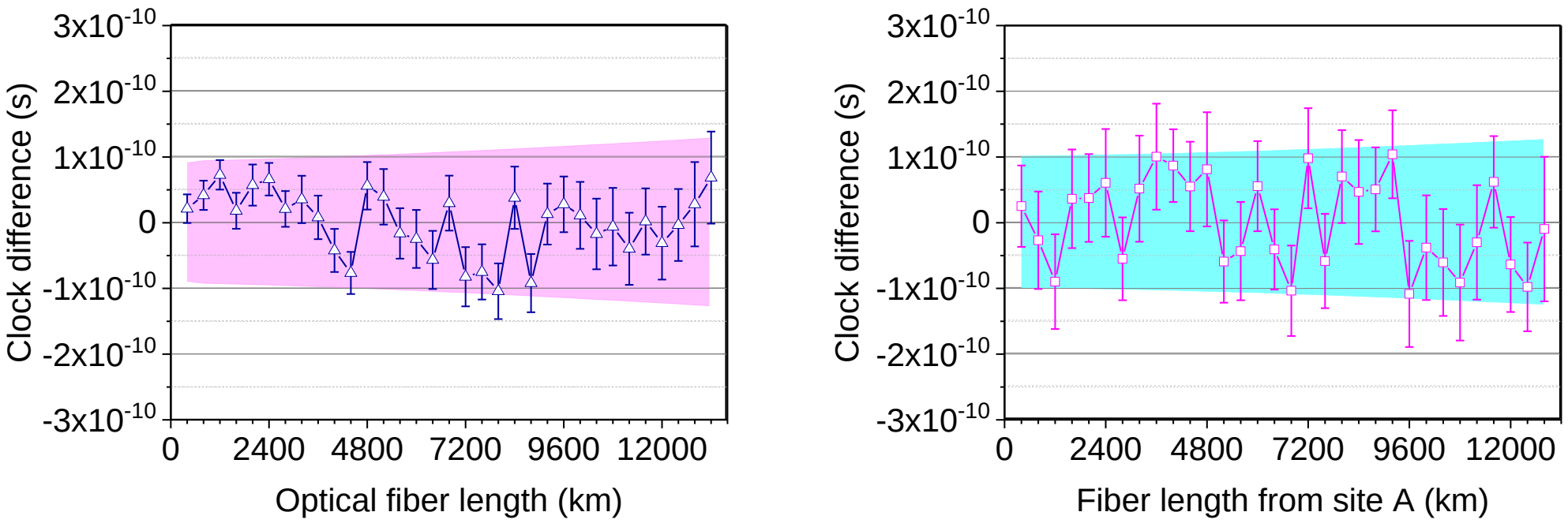

(a) (b)

Fig. 8. Theoretical calculated expanded uncertainty (coverage factor $k$=2) and measured clock differences for (a) different non-calibrated fiber links; (b) different APs along a 13,200 km fiber link.

## 4. Uncertainty Analyses

Besides stability, uncertainty is another essential target for the time transfer system. The background in Fig. 8 shows the calculated expanded uncertainty with the coverage factor of 2 [15]. Since the experimental system is co-located in laboratory, the Sagnac term is excluded. To validate the calculation, the asymmetries of sending and receiving delays for all modems used are calibrated in a common clock configuration [15]. Fig. 8 also presents the measured clock differences over various fiber lengths. As the main link extends from 400 km to 13,200 km, the standard combined uncertainty in theory is not beyond 67 ps and the measured clock differences do not exceed 105 ps. For the APs along a 13,200 km fiber link, the theoretical standard combined uncertainty is less than 66 ps and the measured clock differences are no larger than 109 ps. One can see that the measurements are in well agreement with the theoretical ones. It should be noted that the fiber spans connecting each repeater station is equal in length, the wavelength differences between the repeater stations have neglected effect on the results. The clock difference variation mostly comes from the wavelength difference of adopted SFP transceivers, the receiving optical power-dependent delay of SFP transceivers, and the uncertainty of SR620 etc.

## 5. Conclusion

In summary, we proposed and implemented an OEO repeater station for BTDM-SFSW based time transfer. And multi-access time transfer is successfully demonstrated over the fiber links up to 13,200 km in the laboratory environment. The time deviations of 80 ps/s and 11 ps/$10^5$s are reached, respectively. The calculated combined uncertainty is not beyond 70 ps without the requirement of fiber link calibration. Over the established experimental testbed, the calculated results are validated as well. The proposed scheme opens up the possibility to conduct a high-precision multi-access time transfer over long-haul commercial fiber transmission links.